%% file: howbig.tex
\documentstyle[mncite,psfig]{mn2e}

\newcommand\e{{\mbox{e$^-$}}}
\newcommand\nel{{\mbox{$n_{{\rm e}^-}$}}}

\newcommand\h{{\mbox{H}}}
\newcommand\nh{{\mbox{$n_{\rm H}$}}}
\newcommand\hplus{{\mbox{H$^+$}}}
\newcommand\nhplus{{\mbox{$n_{{\rm H}^+}$}}}
\newcommand\nhtot{{\mbox{$n_{\rm H,tot}$}}}
\newcommand\hminus{{\mbox{H$^-$}}}
\newcommand\nhminus{{\mbox{$n_{{\rm H}^-}$}}}
\newcommand\Rhminus{{\mbox{$R_{{\rm H}^-}$}}}
\newcommand\htwo{{\mbox{H$_2$}}}
\newcommand\nhtwo{{\mbox{$n_{{\rm H}_2}$}}}
\newcommand\fhtwores{{\mbox{$f_{{\rm H}_{2,{\rm res}}}$}}}
\newcommand\htwoplus{{\mbox{H$_2^+$}}}
\newcommand\nhtwoplus{{\mbox{$n_{{\rm H}_2^+}$}}}
\newcommand\Rhtwoplus{{\mbox{$R_{{\rm H}_2^+}$}}}
\newcommand\he{{\mbox{He}}}
\newcommand\heplus{{\mbox{He$^+$}}}
\newcommand\heplusplus{{\mbox{He$^{++}$}}}
\newcommand\nhe{{\mbox{$n_{\rm He}$}}}
\newcommand\nheplus{{\mbox{$n_{\rm He^+}$}}}
\newcommand\nheplusplus{{\mbox{$n_{\rm He^{++}}$}}}
\newcommand\nhetot{{\mbox{$n_{\rm He,tot}$}}}

\newcommand\LCDM{{\mbox{$\Lambda$CDM}}}

\newcommand\SCDM{{\mbox{SCDM}}}
\newcommand\TCDM{{\mbox{$\tau$CDM}}}
\newcommand\Tvir{{\mbox{$T_{\rm vir}$}}}
\newcommand\Tsf{{\mbox{$T_{0.75}$}}}
\newcommand\rvir{{\mbox{$r_{\rm vir}$}}}
\newcommand\tvir{{\mbox{$t_{\rm vir}$}}}
\newcommand\zvir{{\mbox{$z_{\rm vir}$}}}
\newcommand\tdyn{{\mbox{$t_{\rm dyn}$}}}
\newcommand\Mtot{{\mbox{$M_{\rm tot}$}}}

\newcommand\etal{et al.}
\newcommand\Msun{{\mbox{\,M$_\odot$}}}
\newcommand\gta{\,\lower.6ex\hbox{$\buildrel >\over \sim$} \, }
\newcommand\lta{\,\lower.6ex\hbox{$\buildrel <\over \sim$} \, }

\title{How big were the first cosmological objects?}

\author[R. M. Hutchings, \etal]
       {Roger M. Hutchings$^1$, 
        F. Santoro$^1$,
        P. A. Thomas$^1$\thanks{p.a.thomas@sussex.ac.uk}
	\& H. M. P. Couchman$^2$\\ 
{}$^1$ Astronomy Centre, CPES, University of Sussex, Falmer, 
  Brighton, BN1\,9QJ\\
{}$^2$ Department of Physics \& Astronomy, McMaster University,
Hamilton, Ontario, L8S\,4M1, Canada
}

\date{}

\pagerange{\pageref{firstpage}--\pageref{lastpage}}
\pubyear{2000}

\begin{document}

\maketitle

\label{firstpage}

\begin{abstract}
We calculate the cooling times at constant density for halos with
virial temperatures from $100$\,K to $1\times10^{5}$\,K that originate
from a $3\sigma$ fluctuation of a CDM power spectrum in three
different cosmologies.  Our intention is to determine the first
objects that can cool to low temperatures, but not to follow their
dynamical evolution.  We identify two generations of halos: those with
low virial temperatures, $\Tvir\lta9000$\,K that remain largely
neutral, and those with larger virial temperatures that become
ionized.  The lower-temperature, lower-mass halos are the first to
cool to 75 percent of their virial temperature.  The precise
temperature and mass of the first objects are dependent upon the
molecular hydrogen (\htwo) cooling function and the cosmological
model.  The higher-mass halos collapse later but, in this paradigm,
cool much more efficiently once they have done so, first via
electronic transitions and then via molecular cooling: in fact, a
greater residual ionization once the halos cool below 9000\,K results
in an enhanced \htwo\ production and hence a higher cooling rate at
low temperatures than for the lower-mass halos, so that within our
constant-density model it is the former that are the first to cool to
really low temperatures.
We discuss the possible significance of this result in the context of
CDM models in which the shallow slope of the initial fluctuation
spectrum on small scales leads to a wide range of halo masses (of
differing overdensities) collapsing over a small redshift interval.
This ``crosstalk'' is sufficiently important that both high- and
low-mass halos collapse during the lifetimes of the massive stars
which may be formed at these epochs. Further investigation is thus
required to determine which generation of halos plays the dominant
role in early structure formation.
\end{abstract}


\input intro

\input chemistry
\input method
\input results
\input discuss
\input conc

\section{Acknowledgements}
PAT is a PPARC Lecturer Fellow. HMPC thanks the Canadian
Institute for Advanced Research. We are grateful to Todd Fuller
for helpful discussions. FS thanks his parents for their support.

\input refs
\input appendix

\label{lastpage}

\end{document}

%% file: intro.tex
\section{Introduction}
\label{sec:intro}

The study of the first generation of objects in the Universe that are
able to cool sufficiently to collapse and form luminous objects is a
well-defined problem. By definition, there are no stars or other
sources of ionizing radiation, and one does not have to consider
feedback from supernovae and enrichment of the Universe with
metals. The first objects to form arise from the collapse of
high-sigma fluctuations in the background density field.  These peaks
will virialize and begin to cool. Objects with virial temperatures
$T\lta$9\,000\,K are cooled by \htwo\ molecules: the molecules are
excited by collisions with neutral hydrogen and then spontaneously
de-excite with the emission of a photon. In the absence of metals the
dominant cooling mechanisms for 9\,000\,K$\lta T \lta$ 50\,000\,K are
collisional excitation of neutral hydrogen and recombination of
ionized hydrogen, and for $T\gta$50\,000\,K is collisional excitation
of ionized helium.

The question of what are the masses of the first objects to form in a
standard CDM scenario was studied by Tegmark et~al.\ (1997, hereafter
T97) in a landmark paper entitled ``How small were the first
cosmological objects?''  They analytically tracked a top-hat collapse
to the point of virialization, at which point the gas was cooled at
constant density.  They accepted an object as having cooled if it met
the criterion $T(0.75\zvir)\leq0.75\,\Tvir$, where \Tvir\ is the
virial temperature and \zvir\ the virialization redshift.  They found
that the first generation of objects that cooled in a standard CDM
scenario, virialized at a redshift of 27 and had a baryonic mass of
about $10^5$\Msun.  In a later paper, Abel \etal\ (1998) redid the
calculation with a different \htwo\ cooling function and estimated a
very similar virialization redshift but a smaller baryonic mass,
$7\times10^3$\Msun.

In the present paper we adopt a similar approach to T97 but consider
the collapse of not only small objects with virial temperatures
$\leq$9\,000\,K, but also objects of virial temperatures up to
100\,000\,K. The motivation for this is that the high virial
temperatures will partially ionize the gas. Since the gas will then
cool more rapidly than it can recombine, the ionization level at
temperatures $\leq$9\,000\,K will be greater than it otherwise would
have been if no re-ionization had taken place. This in turn
accelerates production of \htwo, ultimately resulting in enhanced
cooling at lower temperatures.  A similar effect was noted by MacLow
\& Shull (1986) and Shapiro \& Kang (1987) for the cooling of gas
behind intergalactic shocks.

We confirm the earlier results on the masses of the \emph{smallest}
halos that can cool, but show that these are not the \emph{first}
objects to do so in this constant-density model.  A plot of cooling
redshift versus halo mass shows two separate maxima corresponding to
halos with virial temperatures of about 4\,000\,K and 11\,000\,K.
Thus there are two distinct generations of primordial halos depending
upon whether or not their virial temperature is high enough to ionize
the gas.  Given our simplified model, each of these generations has a
unique halo mass that will be able to cool first to low temperatures
and form stars.  If they virialized at the same time, then the
higher-mass halos have the potential for more efficient cooling and,
depending upon how the collapse occurs, could win out; however, in
CDM cosmologies, the collapse of larger objects occurs slightly later
than the collapse of smaller ones at the same overdensity relative to
the rms at each scale.  The enumeration of the difference in cooling
times between the two generations of halos is the main topic of this
paper.  We show that this difference is of order 10\,Myr, comparable
to the lifetimes of massive stars. It is thus possible that the second
generation of massive halos could collapse before
feedback from the first stars affects their internal composition.  The
result of this paper, therefore, is to argue that more detailed
modelling is required to establish the relative importance of the two
generations for subsequent structure formation.

A second difference between our study and previous ones is that we
consider more up-to-date cosmological models.  In particular, we
choose a density fluctuation spectrum with less power on small
scales.  In a critical density universe, this has the effect of
delaying collapse until a redshift of 11.  Higher redshifts can be
recovered by the introduction of a cosmological constant.

We describe our chemical model in Section~\ref{sec:chemistry} and our
numerical method in Section~\ref{sec:method}.  The results are
presented in Section~\ref{sec:results} and discussed in
Section~\ref{sec:discuss}.  Finally, we summarize our conclusions in
Section~\ref{sec:conclusions}.

%

%
%

%% file: chemistry.tex
\section{Gas Chemistry and Cooling Functions}
\label{sec:chemistry}

In this section we present a minimal model of the gas chemistry needed
to accurately follow the temperature evolution of the gas.  Our model
is closely based on those of Abel et~al.\ (1997) and Fuller \& Couchman
(2000) who both did a thorough search to identify the key reactions, but
differs from theirs in the manner in which it handles the 
production and destruction of \hminus\ and \htwoplus.  As a
consequence, we end up with slightly different terms in our final
equation for \htwo\ production.  In principle this could be quite
important; however it seems to make little difference to the results
over the parameter ranges considered in this paper.

Table~\ref{hmtab} lists the important reactions for the combinations
of temperature and species abundances that we consider.  The model is
applicable to halos with temperatures between 100\,K and 100\,000\,K
and redshifts up to 50.  Note that Hydrogen and Helium do not interact
chemically.  The latter is included in order that the ionization level
be correctly reproduced at high temperatures, but its omission would
make very little difference to the results presented in this paper.
We do not include photo-ionization from cosmic microwave background
photons which is only important at redshifts in excess of 100.

\begin{table*}
\caption{This table summarizes the important reactions needed in order
to calculate accurately the abundance of \htwo. References are: HTL
Haiman, Thoul \& Loeb (1996); GP Galli \& Palla (1998); FC Fuller \&
Couchman (2000); SLD Stancil, Lepp \& Dalgarno (1998); AAZN Abel,
Anninos, Zhang \& Norman (1997).}
\label{hmtab}
\begin{tabular}{llll}
\emph{ } & Reaction& Rate/cm$^3$s$^{-1}$&
Reference \\[2pt]
\hline 

 1 & $\h + \e \longmapsto \hplus + 2\e$
   & $5.9\times10^{-11}\,T_0^{0.5}(1+T_5^{0.5})^{-1}\exp(-1.58/T_5)$
   & HTL \\[2pt]

 2 & $\hplus + \e \longmapsto \h + \gamma$ 
   & $3.3\times10^{-10}\,T_0^{-0.7}(1+T_6^{0.7})^{-1}$
   & HTL \\[2pt]

 3 & $\h + \e \longmapsto \hminus + \gamma$ 
   & $1.4\times 10^{-18}\,T_0^{0.93}\exp(-T_4/1.62)$
   & GP \\[2pt]


 4 & $\hminus + \h \longmapsto \htwo + \e$
   & $1.3\times10^{-9}$
   & FC \\[2pt]

 5 & $\hminus +\hplus \longmapsto 2\h $
   & $4.0\times 10^{-6}\ T_0^{-0.5}$
   & FC\\[2pt]

 6 & $\h + \hplus \longmapsto \htwoplus + \gamma$
   & $2.1\times10^{-23}\,T_0^{1.8}\exp(-2/T_1)$
   & SLD\\[2pt]


 7 & $\htwoplus + \h \longmapsto \htwo + \hplus$
   & $6.4\times 10^{-10}$
   & GP \\[2pt]

 8 & $\htwoplus + \e \longmapsto 2\h $
   & $1.2\times 10^{-7}\ T_0^{-0.4}$
   & SLD \\[2pt]

 9 & $\htwo + \hplus \longmapsto \htwoplus +\h$
   & $\min\big(3.0\times 10^{-10}\exp(-2.11/T_4)$,
     $1.5\times 10^{-10}\exp(-1.40/T_4)\big)$
   & GP \\[2pt]

10 & $\htwo + \h \longmapsto 3\h$
   & $7.1\times 10^{-19}\,T_0^{2.01}(1+2.13\,T_5)^{-3.51}
     \exp\left(-5.18/T_4\right)$
   & AAZN \\[2PT]

11 & $\htwo + \e \longmapsto 2\h + \e$
   & $4.4\times 10^{-10}\,T_0^{0.35}\exp(-1.02/T_5)$
   & GP \\[2pt]

12 & $\he + \e \longmapsto \heplus + 2\e$
   & $2.4\times 10^{-11}\,T_0^{0.5}\,(1+T_5^{0.5})^{-1}\exp(-2.85/T_5)$ 
   & HTL \\[2PT]

13 & $\heplus + \e \longmapsto \he + \gamma$
   & $ 1.5\times10^{-10}\,T_0^{-0.64}
     +1.9\times10^{-3}\,T_0^{-1.5}\exp(-5.64/T_5)\,(0.3+\exp(9.40/T_4))$ 
   & AAZN \\[2PT]

14 & $\heplus + \e \longmapsto \heplusplus + 2\e$
   & $5.7\times10^{-12}\,T_0^{0.5}\,(1+T_5^{0.5})^{-1}\exp(-6.32/T_5)$ 
   & HTL \\[2PT]

15 & $\heplusplus + \e \longmapsto \heplus + \gamma$
   & $1.3\times10^{-9}\,T_0^{-0.7}\,(1+T_6^{0.7})^{-1}$
   & HTL

\end{tabular}
\end{table*}

\subsection{\htwo\ production and destruction}
\label{sec:h2chem}

The equation for the rate of change of \htwo\ abundance is derived in
the appendix:
\begin{eqnarray}
\label{eq:htwolong}
\lefteqn{ \frac{d\nhtwo }{dt} = \nh ^2\left[
\nel {R_3R_4\over \Rhminus } + \nhplus {R_6R_7\over \Rhtwoplus }
\right] } \\
& & {} - \nhtwo 
\left[ \nhplus\nel{R_9 R_8\over\Rhtwoplus} + \nh R_{10} + \nel R_{11} 
\right], \nonumber
\end{eqnarray} 
where 
\begin{equation}
\Rhminus =\nh R_4+\nhplus R_5,
\end{equation}
\begin{equation}
\Rhtwoplus =\nh R_7+\nel R_8,
\label{eq:htwoplus}
\end{equation}
and $R_n$ is the rate of Reaction n from Table~\ref{hmtab}.

We next discuss the regimes under which each of these terms is important.

\subsubsection{Formation of \htwo}
The \hminus\ channel is the dominant \htwo\ formation path.  The
\htwoplus\ channel is important only at high redshifts ($z\gta200$
when the \hminus\ channel is suppressed by photo-destruction of
\hminus that we have omitted here) or high temperatures.  However, at
high temperatures, the \htwo\ is rapidly destroyed and for all the
models that we consider in this paper, omitting the \htwoplus\
formation channel makes no difference to the final abundance of \htwo.

\subsubsection{Destruction of \htwo}
It is the process of \htwo\ destruction where the present work differs
slightly from that of Abel \etal\ (1997) and Fuller \& Couchman (2000). 
In these two papers, the term for destruction by \hplus\ is given as
\begin{equation}
\nhtwo\nhplus R_9,
\end{equation} 
whereas we have
\begin{equation}
\nhtwo\nhplus R_9{\nel R_8\over\Rhtwoplus} 
=\nhtwo\nhplus R_9
\left(1-{\nh R_7\over \Rhtwoplus }\right).
\end{equation}
This takes into account that, at low ionization levels,
much of the \htwoplus\ that is produced will be immediately converted
back to \htwo\ so that there will be no net destruction.  In some of
the models that we consider, this can make transitory differences of a
factor of 10 or more in the \htwo\ abundance.  However, as \htwo\
production takes over from destruction at low temperatures, the effect
on the final \htwo\ abundance is quite small (at most a few percent).

With the correct rate for the \htwoplus\ destruction channel, there is
a small parameter range for which H10 becomes the dominant destruction
process.  However, this is so fleeting that it makes a negligible
difference to the results and can safely be omitted.

We note that the previous papers consider only low temperatures,
$T<6\,000\,$K for which \htwo\ destruction is relatively unimportant
compared to its production, and the error in their results is
negligible.  Nevertheless, a minimal model that includes \htwo\
destruction should use our Equation~\ref{eq:htwolong}.

\subsection{Ionization level}
\label{sec:ionization}

In principle, the Equation for the rate of change of \hplus\ abundance
is every bit as complicated as that for \htwo.  However, it turns out
in practice that there are only two important terms:
\begin{equation}
\frac{d\nhplus }{dt} = \nel\nh R_1 - \nel\nhplus R_2.
\label{eq:rec}
\end{equation}
Similarly, for the Helium species,
\begin{equation}
\frac{d\nhe}{dt} = \nel\nheplus R_{13} - \nel\nhe R_{12},
\label{eq:rhe}
\end{equation}
\begin{equation}
\frac{d\nheplusplus }{dt} = \nel\nheplus R_{14} - \nel\nheplusplus R_{15}.
\label{eq:rheplusplus}
\end{equation}
The number density of electrons is given by charge conservation:
\begin{equation}
\nel = \nhplus + \nheplus + 2\nheplusplus.
\label{eq:elec}
\end{equation}

\subsection{Cooling Terms}

At low temperatures the main coolant is molecular hydrogen.  We use
the cooling rate given in Galli \& Palla (1998)
as summarised by Fuller \& Couchman (2000).  This gives a fit to the
low-density limit of the calculations of Martin \etal\ (1996) and
Forrey \etal\ (1997) which together cover a wide temperature range:
\[
\log_{10}\left(\Lambda_{\scriptsize\htwo}(T)
\over{\nh\nhtwo\rm erg\,cm^3\,s^{-1}}\right) = -103.0+97.59\,T_{\log}
\]
\begin{equation}
\hspace*{1cm} -48.05\,T_{\log}^2 +10.80\,T_{\log}^3-0.9032\,T_{\log}^4,
\label{eq:htwocool}
\end{equation}
where $T_{\log}=\log_{10}(T/{\rm K})$.

For all other cooling processes, we use the rates given in Haiman,
Thoul \& Loeb (1996).  The most important of these are as follows.
At temperatures, $T\gta10\,000\,K$,
collisional excitation and (less importantly) ionization of atomic
Hydrogen take over from molecular Hydrogen cooling:
\begin{equation}
\frac{\Lambda_{\rm H,ce}(T)}{\nel\nh\rm erg\,cm^3\,s^{-1}}=
7.50\times10^{-19}{1\over 1+T_5^{1\over2}}\,e^{-{1.183\over T_5}}
\end{equation} 
\begin{equation}
\frac{\Lambda_{\rm H,ci}(T)}{\nel\nh\rm erg\,cm^3\,s^{-1}}=
4.02\times10^{-19}{T_5^{1\over2}\over 1+T_5^{1\over2}}
\,e^{-\frac{1.578}{T_5}}
\end{equation} 
where $T_{\rm{n}}$ is the temperature in units of $10^{\rm{n}}$\,K.

At even higher temperatures, $50\,000\,K\lta T\lta100\,000$,
collisional excitation of \heplus\ is the dominant coolant.
\begin{equation}
\frac{\Lambda_{\rm He^+,ce}(T)}{\nel\nheplus\rm erg\,cm^3\,s^{-1}}=
5.54\times10^{-17}\frac{ T_0^{-0.397}}{1 + T_5^{1\over2}}\,e^{-{4.737\over T_5}}.
\end{equation}

We include other processes from Haiman, Thoul \& Loeb (1996) in our model,
including inverse Compton cooling from cosmic microwave background
photons, but at the modest redshifts ($z<50$) and high densities of
the collapsed halos that we consider in this paper none contribute at
more than the few percent level.

%% file: method.tex
\section{Numerical Procedure}
\label{sec:method}

\subsection{General approach}

The approach that we have adopted for this paper is to specify an
initial temperature, ionization level and \htwo\ fraction and to
follow their evolution at constant density, using the relevant cooling
terms and reaction rates discussed in the previous section.  The
assumption of constant density is only valid for halos whose cooling
times are much longer than their dynamical times.  It is adequate for
the purposes of this paper to assess if a given halo at constant
density can begin to cool efficiently, but not to follow the collapse
of halos once they do cool.

The equations are integrated using the RK4 integrator from Press
\etal\ (1992) modified to use an adaptive timestep that allowed neither the
abundances nor the temperature to vary by more than 0.1 per cent
during a timestep.  We also tried a Bulirsch-Stoer integrator which
gave identical results for a slightly poorer performance.  For tests,
we integrated simplified networks of equations for which a solution
can be obtained analytically: for example, Equations~15 and 16 of T97.

The initial halo parameters are picked to represent the conditions of
a virialized object which has collapsed from a $3\sigma$ peak in a CDM
scenario, with virial temperatures, \Tvir\ in the range 100\,K to
100\,000\,K.  We consider two different measures of cooling.  Firstly,
we measure how long it takes halos to cool to $\Tsf=0.75\,\Tvir$---T97
define an object to have cooled if its temperature decreases by 25 per
cent or more in the time that redshift does likewise.  We find two
distinct populations of clouds: low-mass ones that cool via molecular
hydrogen and high-mass ones that cool via electronic transitions.
Secondly, we look at the amount by which an object can cool in one
dynamical time.  Clouds with virial temperatures greater than
9\,000\,K (i.e.\ those that have re-ionized) form later (at the same
$\sigma$) but cool more effectively than lower-mass clouds, and have a
much greater \htwo\ fraction once they have cooled to low
temperatures.

\subsection{Cosmological Models}

We compute the collapse redshift of objects arising from $3\sigma$
peaks of a CDM power spectrum. The spectrum was calculated using a
real-space top-hat window function, the transfer function of Bond \&
Efstathiou (1984) for scales above 1\,$h^{-1}$Mpc and the transfer
function of Bardeen \etal\ (1986), or BBKS, for smaller scales. We
choose this combination as the Bond \& Efstathiou transfer function is
more accurate than BBKS but makes no attempt to accurately calculate
the function on scales below 1\,$h^{-1}$Mpc.

We present results for three different cosmological models, as listed
in Table~\ref{tab:cospar}.  
\begin{table*}
\caption{Cosmological parameters for the three models: model name;
density parameter; cosmological constant in units of
$\lambda_0=\Lambda/3H_0^2$; Hubble parameter in units of
$h=H_0/100$km\,s$^{-1}$Mpc$^{-1}$; root-mean-square dispersion of the
density within spheres of radius 8\,$h^{-1}$Mpc; residual ionization
fraction.}
\label{tab:cospar}
\begin{center}
\begin{tabular}{lcccccccc}
Name& $\Omega_0$& $\lambda_0$& $\Omega_{b0}$& $h$& $\Gamma$&
$\sigma_8$& $n_{e{\rm,res}}$/\nhtot\\
\hline
\SCDM& 1.0& 0.0& 0.076& 0.5& 0.43& 0.60& $1.58\times10^{-4}$\\
\TCDM& 1.0& 0.0& 0.184& 0.5& 0.21& 0.60& $6.52\times10^{-5}$\\
\LCDM& 0.35& 0.65& 0.038& 0.7& 0.21& 0.90& $1.33\times10^{-4}$\\
\end{tabular}
\end{center}
\end{table*}
For comparision with previous work, we first choose the standard, cold
dark matter cosmology (\SCDM) with power spectrum shape parameter,
$\Gamma=0.43$.  When normalised to the COBE results, this power
spectrum is now known to have too much power on small scales and so
the other two models that we consider use a smaller value,
$\Gamma=0.21$.

The \LCDM\ model is our best-guess at the most favourable CDM model.
The combination of $\Omega_0$, $\Omega_{b0}$ and $h$ naturally produces a
power spectrum with the correct value of $\Gamma$, and the Hubble
parameter and baryon fraction both lie close to currently preferred
values (see, for example, Freedman \etal\ 2001 for the former and Ettori \&
Fabian 1999 for the latter).  The normalisation of the power
spectrum, $\sigma_8$, is chosen to reproduce the correct local
abundance of rich clusters (Viana \& Liddle 1996).

The \TCDM\ model is an attempt to salvage a critical density model in
the post-COBE epoch.  It has some motivation in decaying neutrino
scenarios (e.g.\ Hansen \& Villante 2000) but is mostly phenomenological in
nature.  For this reason, we have fixed $\Gamma$ to have the same
value as in the \LCDM\ model, but have used a lower value of
$\sigma_8$ to again reproduce the rich cluster abundance.  In
addition, we have had to take a lower value of the Hubble parameter in
order to overcome the age problem, and a baryon density considerably
higher than that predicted by primordial nucleosynthesis to give the
correct baryon fraction in clusters.

\subsection{Initial halo properties}

We calculate the initial properties of halos by assuming that they
have settled down into virial equilibrium.  This will be a good
approximation for halos whose cooling time substantially exceeds their
dynamical time and should therefore enable us to identify 
objects that are able to cool with reasonable accuracy.  To correctly
model the properties of halos whose cooling time is shorter than their
dynamical time would require a more sophisticated model.

In an isothermal sphere, the density is strongly peaked towards the
centre of the halo.  Here we follow the evolution of gas which has a
density fixed at the mean value within the virialised halo,
\begin{equation}
\rho_{\rm vir}= \left(\Delta_c\over\Omega\right)\rho_{b0}(1+\zvir)^3
\end{equation} 
where \zvir\ is the virialization redshift; $\rho_{b0}$ is the current
mean density of baryons in the Universe; and $\Delta_c$ is the mean
overdensity relative to the critical density within the virialized
halo which we take to be $18\pi^2$ for the \SCDM\ and \TCDM\
cosmologies, and $18\pi^2\Omega^{0.45}$ for the \LCDM\ cosmology,
where $\Omega$ is the density parameter at the time of virialization
(Eke, Navarro \& Frenk, 1998).  We take a hydrogen mass fraction (in
all its forms: neutral, ionized and molecular) of $X=0.755$.  Thus
\begin{equation}
\nhtot\equiv\nh+\nhplus+2\nhtwo=\rho_{\rm vir}X.
\end{equation}
Similarly,
\begin{equation}
\nhetot\equiv\nhe+\nheplus+\nheplusplus=\rho_{\rm vir}Y,
\end{equation}
where $Y=(1-X)=0.245$ is the Helium mass fraction.

For the \TCDM\ and \LCDM\ cosmologies, the residual ionization level
from the early Universe is taken from Peebles (1968, 1993) but divided
by two as an attempt to compensate for the neglect of stimulated
recombination as suggested by the results of Grachev \& Dubrovich
(1991) and Sasaki \& Takahara (1993). Although dividing by two is
somewhat arbitrary, it does result in a more accurate estimate of the
initial ionization level than many previous papers without explicitly
solving the time-dependent evolution of the ionization level with the
inclusion of stimulated recombination.  We take
\begin{equation}
{n_{e{\rm,res}}\over\nhtot}=
6\times 10^{-6}\Omega_0^{1\over2}(\Omega_{b0}h)^{-1},
\label{eq:feres}
\end{equation} 
which gives the residual ionization levels listed in Table~\ref{tab:cospar}.

For halos hotter than about 8\,000--9\,000\,K, the gas will be ionized
above the residual value by the process of virialization.  For these
we use the equilibrium values determined by solving
Equations~\ref{eq:rec}, \ref{eq:rhe} and \ref{eq:rheplusplus}
for d\nhplus/d$t=$d\nhe/d$t=$d\nheplusplus/d$t=0$:
\begin{equation}
{\nhplus\over\nhtot} ={R_1\over R_1+R_2},
\label{eq:fhpluseq}
\end{equation}
\begin{equation}
{\nheplus\over\nhetot}={R_{12}R_{15}
\over R_{13}R_{15}+R_{12}R_{14}+R_{12}R_{15}}
\label{eq:fheeq}
\end{equation}
and
\begin{equation}
{\nheplusplus\over\nhetot}={R_{12}R_{14}
\over R_{13}R_{15}+R_{12}R_{14}+R_{12}R_{15}}.
\label{eq:fhepluspluseq}
\end{equation}

The background level of \htwo\ in the Universe is taken to be
$\fhtwores=1.1\times10^{-6}$ as calculated by Galli \& Palla (1998).
    
We take the initial temperature to be
\begin{eqnarray}
\Tvir &=& {\mu m_H\over k_B}{G\Mtot\over2\rvir} \\
&\approx& 40.8\,{\mu\over 1.225}\,(1+z_{\rm vir})
\left(\Delta_c h^2\over18\pi^2\Omega_0\right)^{1\over3}
\left(\Mtot\over10^5\Msun\right)^{2\over3}{\rm K}.\nonumber
\end{eqnarray} 
Here \Mtot\ is the total mass (dark plus baryonic) which we assume is
distributed as an isothermal sphere within the virial radius, \rvir;
$m_H$ is the mass of a hydrogen atom; $k_B$ is the
Boltzmann constant; G is the gravitational constant; and
$\mu$ is the mean mass of particles in units of $m_H$:
\begin{equation}
\mu={\nhtot+4\nhetot\over\nhtot-\nhtwo+\nhetot+\nel}.
\label{eq:mu}
\end{equation}
When the molecular Hydrogen abundance is low, then
$\mu$ varies between 1.23 at low temperatures, $T\lta$10\,000\,K, and
0.59 at very high temperatures, $T\gta$100\,000\,K.

\begin{figure}
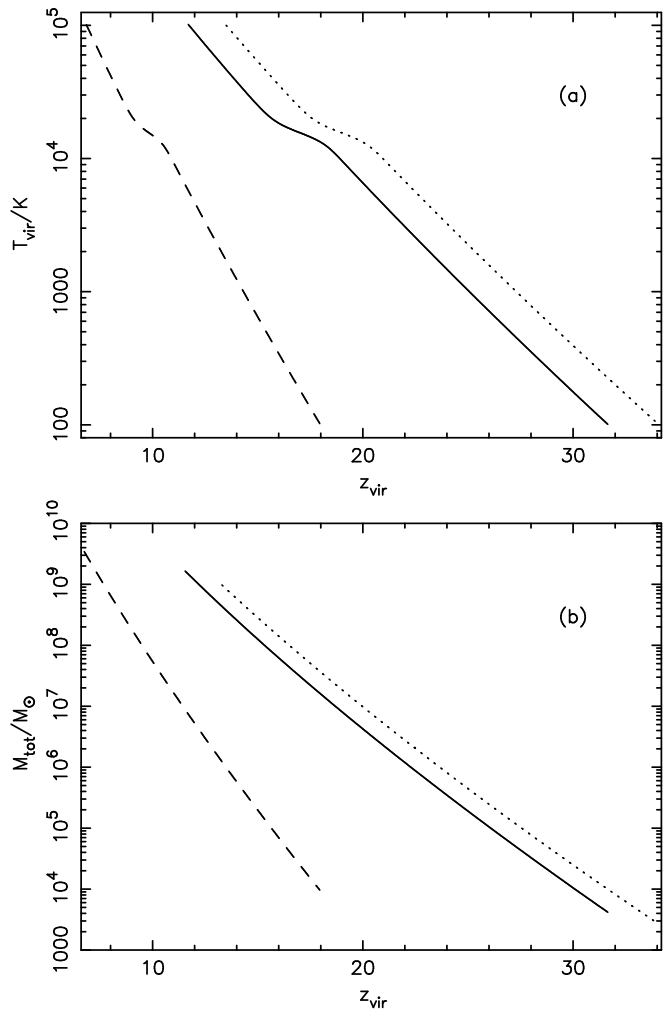

\begin{center}
\psfig{width=8.7cm,angle=270,file=tvir.ps}
\psfig{width=8.7cm,angle=270,file=mvir.ps}
\caption{This figure shows the relation between virialization redshift
and (a) temperature and (b) total mass for the three cosmologies:
\SCDM\ (solid line), \TCDM\ (dashed line), and \LCDM (dotted line).}
\label{fig:tmvir}
\end{center}
\end{figure}

We shall use the halo virialization redshift, \zvir, as our ordinate
in most of the plots that follow.  This can be converted to
temperature or mass in a cosmology-dependent way.  We plot these
relations in Figure~\ref{fig:tmvir} and list examples of the
conversion between temperature and other quantities in
Table~\ref{tab:tzm}.  The kink in the temperature profiles corresponds
to the changing value of $\mu$ from Equation~\ref{eq:mu} during
Hydrogen ionization (there is a second kink due to ionization of
Helium but that is barely discernible).  Note that
halos of a given temperature have similar total masses in each
cosmology, although they may virialize at very different redshifts.

\begin{table}
\caption{Conversion between virialization temperature, redshift and
mass for the three models: temperature; model; virialization redshift;
total mass; baryonic mass.}
\label{tab:tzm}
\begin{center}
\begin{tabular}{rrccc}
$T/$K& Model& $z_{\rm vir}$& $M_{\rm tot}/\Msun$& $M_{\rm bary}/\Msun$\\
\hline
    100& \SCDM& 31.6& $4.2\times10^3$& $3.2\times10^2$\\
       & \TCDM& 18.0& $9.2\times10^3$& $1.7\times10^3$\\
       & \LCDM& 34.0& $2.7\times10^3$& $2.9\times10^2$\\
 1\,000& \SCDM& 25.0& $1.8\times10^5$& $1.4\times10^4$\\
       & \TCDM& 14.3& $4.1\times10^5$& $7.5\times10^4$\\
       & \LCDM& 27.3& $1.1\times10^5$& $1.2\times10^4$\\
10\,000& \SCDM& 18.9& $8.7\times10^6$& $6.6\times10^5$\\
       & \TCDM& 10.9& $1.9\times10^7$& $3.5\times10^6$\\
       & \LCDM& 21.0& $5.4\times10^6$& $5.9\times10^5$\\
100\,000& \SCDM& 11.6& $1.6\times10^9$& $1.2\times10^8$\\
        & \TCDM&  6.8& $3.3\times10^9$& $6.1\times10^8$\\
        & \LCDM& 13.3& $9.7\times10^8$& $1.1\times10^8$\\
\end{tabular}
\end{center}
\end{table}

%% file: results.tex
\section{Results}
\label{sec:results}

\subsection{\SCDM}
\label{sec:scdm}

To compare with previous work, we first consider the collapse of halos
in the \SCDM\ cosmology.  We do this for two different forms of the
\htwo\ cooling function.  The first, described above in
Equation~\ref{eq:htwocool}, is from Galli \& Palla (1998, hereafter
GP98); the second is from Lepp \& Shull (1984; hereafter LS84).  This
latter form is the one used by Abel \etal\ (1998); it gives an order
of magnitude more cooling at temperatures below 1000\,K and so favours
the collapse of low-mass, low-virial temperature objects.

\begin{figure}
\begin{center}
\psfig{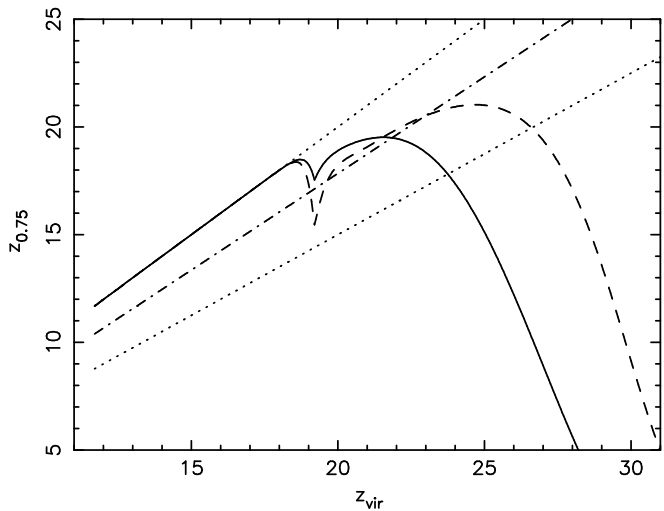}
\caption{Redshift at which a gas halo will cool to 75 per cent of its
initial temperature, $z_{0.75}$, versus virialization redshift, \zvir,
for cosmology \SCDM\ for two different cooling functions: dashed curve,
LS84; solid curve, GP98.  The upper and lower dotted lines show the
relations $z_{0.75}=z_{\rm vir}$ and $z_{0.75}=0.75\,z_{\rm vir}$,
respectively; the dash-dotted line shows the redshift one dynamical
time after virialization.}
\label{fig:scdm1}
\end{center}
\end{figure}

Initially the gas within the halos is assumed to be shocked to the
virial temperature and to be pressure supported.  If it can cool
significantly within one dynamical time then it will contract towards
the centre of the halo in order to maintain pressure support.
Accordingly, we plot in Figure~\ref{fig:scdm1} the redshift,
$z_{0.75}$, at which the gas (if it maintains constant density) will
lose 25 percent of its initial energy.  The solid curve was generated
using the GP98 cooling function and the dashed curve using that of
LS84.  The lower dotted line shows the relation $z_{0.75}=0.75\,z_{\rm
vir}$ which is the condition used by T97 to separate clouds which can
cool from those which cannot.  The dashed line shows the redshift one
dynamical time, \tdyn, after the time of virialization, \tvir, where
we set $\tdyn=\tvir/4\surd2$.

First concentrate on the GP98 curve.  This intersects the line
$z_{0.75}=0.75\,z_{\rm vir}$ at $z_{\rm vir}\approx23.8$ which means
that the smallest clouds that can cool according to the T97 criterion
will have virial temperatures of \Tvir=1\,600\,K and masses of
$4\times10^5$\Msun.  The most important quantity for comparison with
the results of T97 is the temperature which agrees very well with that
given in their Figure~5.  For a virialization redshift of 23.8, they
require a slightly higher temperature of 2\,500\,K for collapse, but
this difference is attributable to our having a slightly higher
baryon density and \htwo\ cooling rate.

T97 assume, within the constraints of the constant density model, that the smallest halos that can cool will be the first to
do so, but that is not the case.  As one moves to lower \zvir\ and
higher \Tvir, the efficiency of \htwo\ production increases and the
cooling time decreases.  Consequently, the first objects to cool to 75
per cent of their initial temperature virialise later at $z_{\rm
vir}\approx21.6$, and have higher virial temperatures (3\,700\,K) and
masses ($1.6\times 10^6$\Msun).  This maximum in $z_{0.75}$
corresponds roughly to the redshift one dynamical time after
virialisation.

At higher temperatures the efficiency of \htwo\ cooling continues to
increase but cannot compensate for the later virialization redshift and
so $z_{0.75}$ decreases once more.  Note the cusp in the curve at
$\zvir\approx19.2$, corresponding to $\Tvir\approx 9\,000\,K$.  As this
temperature is approached, \htwo\ starts to be destroyed and so its
abundance plummets and the cooling time increases.  However, at higher
temperatures collisional excitation and ionization take over and these
are much more efficient cooling processes so the cooling time drops
once more.  In fact for temperatures above 11\,000\,K the cooling is
essentially instantaneous (i.e.~the cooling time much less than the
dynamical time).

Thus the two peaks in Figure~\ref{fig:scdm1} correspond to two
different classes of object: the first to collapse, at
$z_{0.75}=19.4$, are dominated by molecular cooling and the second, at
$z_{0.75}=18.5$ by electronic transitions.  As we are interested in
the first objects to cool, it might be thought that the smaller halos
are the more important, but this is not entirely clear because the
T97 criterion does not trace the cooling down to very low
temperatures.  In fact the more massive halos have a large residual
ionization which results in a greater \htwo\ production once they have
cooled below 9\,000\,K, and so they may be the first objects to cool to
really low temperatures.  We will discuss this further in
Section~\ref{sec:tcdm}, below.

Turning now to the LS84 curve, we see that this is qualitatively
similar.  However, it predicts shorter cooling times for low-mass
clouds because the \htwo\ cooling rate is much higher.  Hence the
smallest objects that can collapse have a lower mass and temperature
than for the GP98 cooling function.  The lowest-mass objects that
satisfy the T97 criterion virialize at a redshift of 26.6 and have
virial temperatures of 570\,K---this seems to agree reasonably well
with the predictions from Figure~12 of Abel \etal\ (1998).  The peak
of the curve has moved to $z_{\rm vir}\approx24.7$ and the temperature
and mass of these first objects are significantly lower, 1\,100\,K and
$2.3\times10^5\Msun$, respectively.  By contrast, the properties of
the halos corresponding to the higher-mass peak are only slightly
modified.

We use the more up-to-date GP98 cooling function throughout the
rest of this paper.

\subsection{\TCDM}
\label{sec:tcdm}

\begin{figure}
\begin{center}
\psfig{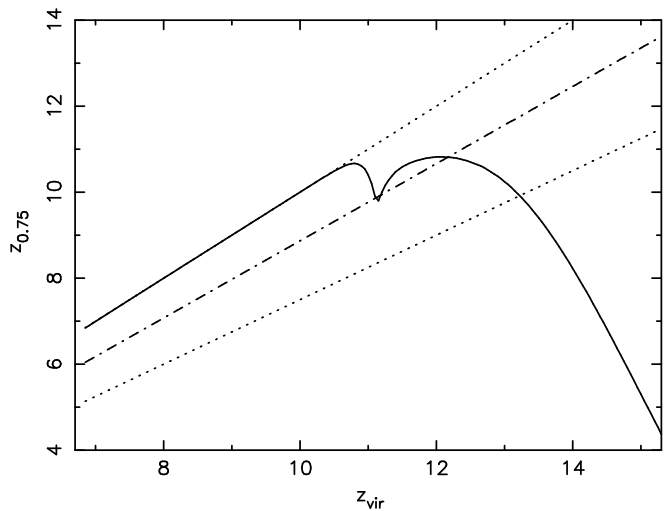}
\caption{The solid line shows the redshift at which a gas halo will
cool to 75 per cent of its initial temperature, $z_{0.75}$, versus
virialization redshift, \zvir, for cosmology \TCDM\ using the GP98
cooling function.  The upper and lower dotted lines show the
relations $z_{0.75}=z_{\rm vir}$ and $z_{0.75}=0.75\,z_{\rm vir}$,
respectively; the dash-dotted line shows the redshift one dynamical
time after virialization.}
\label{fig:tcdm1}
\end{center}
\end{figure}

We now turn to the \TCDM\ cosmology.  Like the \SCDM\ cosmology, this
has a critical density of matter and a high baryon density, but it has
a spectral shape that gives less power on small scales and hence the
first objects collapse at much lower redshift.  This is shown in
Figure~\ref{fig:tcdm1} which is the analogue of Figure~\ref{fig:scdm1}
but for the \TCDM\ cosmology.  We can see that gas halos first manage
to cool to 0.75\,\Tvir\ only at redshift $z_{0.75}=10.8$: these have
virial temperatures of 4\,500\,K and total masses of $5.0\times10^6$\Msun.

\begin{figure}
\begin{center}
\psfig{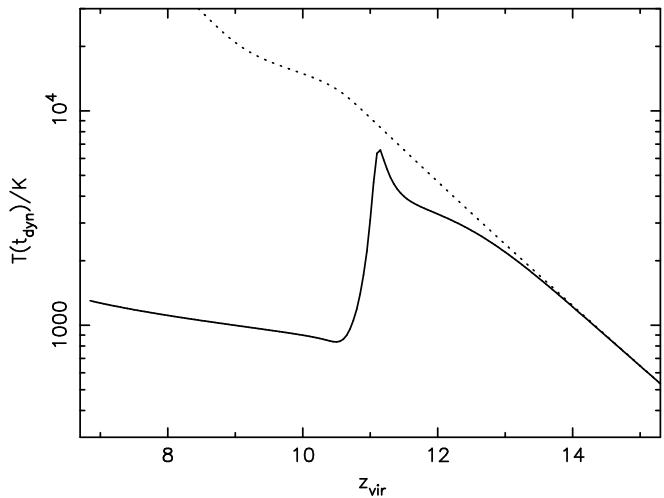}
\caption{Temperature obtained after one dynamical time versus
virialization redshift, for cosmology \TCDM.  The dotted line shows
the initial, virial temperature of the halo.  The cusp in the curve
corresponds to a virial temperature of 9\,000\,K.} 
\label{fig:tcdm2}
\end{center}
\end{figure}

In order to form stars, the gas has to cool to much less than \Tsf.
Therefore we consider a second measure of cooling:
Figure~\ref{fig:tcdm2} shows the final temperature of the halo after
one dynamical time (once again assuming constant density---this will
be a good assumption only for clouds which cool by a small amount in a
dynamical time; for others it will underestimate their cooling rate).
The coldest clouds are those to the right-hand-side of the Figure, but
this is only because they were born with low temperatures: they have
cooled very little.  The final temperature is an increasing function
of virial temperature up to about 9\,000\,K when it suddenly plummets,
so that somewhat surprisingly, the larger, higher virial temperature
halos, have cooled to a lower temperature than the smaller ones.
\begin{figure}
\begin{center}
\psfig{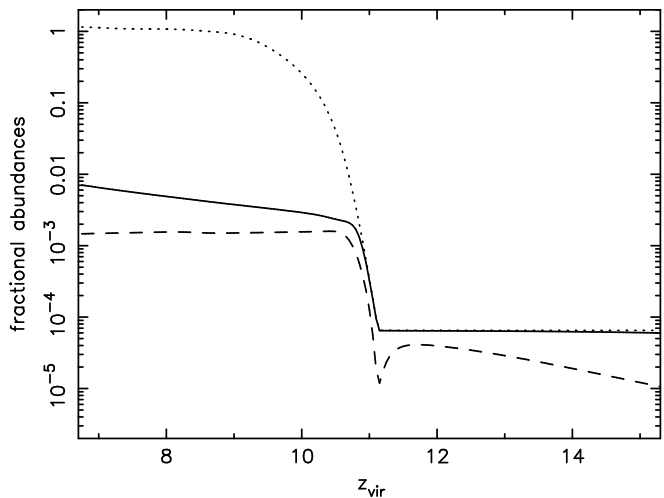}
\caption{Ionization fraction (solid line) and \htwo\ abundance (dashed
line) after one dynamical time, for cosmology \TCDM.  The dotted line
shows the initial ionization level of the halo.} 
\label{fig:tcdm3}
\end{center}
\end{figure}
The reason for this is shown in Figure~\ref{fig:tcdm3} which shows that
the \htwo\ abundance after one dynamical time is larger in high-mass
halos than low-mass ones.  This is because they become highly ionized
and their residual ionization once they have cooled back down below
9\,000\,K is greater than that of low-mass halos, which in turn
results in a greater production of \htwo.  Note that the cooling rate
of these high-mass halos will continue to be greater than that of
low-mass halos even after one dynamical time.  It is therefore
conceivable that they will be the first to cool to really low
temperatures.  Because the gas is free to move around on this
timescale, however, realistic dynamical simulations are required to
model the problem accurately.

\subsection{\LCDM}
\label{sec:lcdm}

\begin{figure}
\begin{center}
\psfig{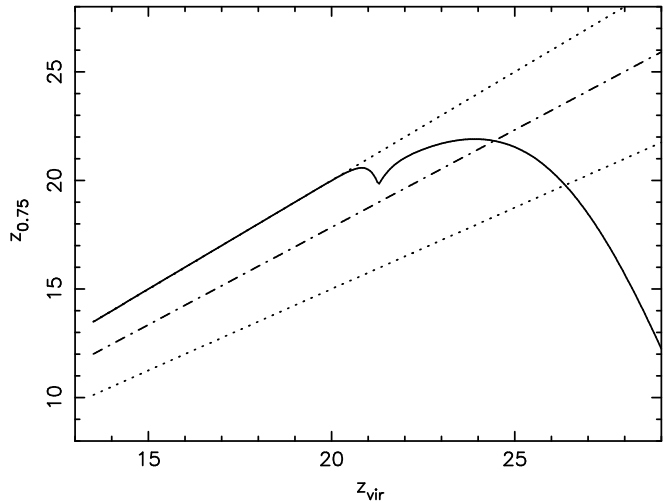}
\caption{The solid line shows the redshift at which a gas halo will
cool to 75 per cent of its initial temperature, $z_{0.75}$, versus
virialization redshift, \zvir, for cosmology \LCDM\ using the GP98
cooling function.  The upper and lower dotted lines show the
relations $z_{0.75}=z_{\rm vir}$ and $z_{0.75}=0.75\,z_{\rm vir}$,
respectively; the dash-dotted line shows the redshift one dynamical
time after virialization.}
\label{fig:lcdm1}
\end{center}
\end{figure}

\begin{figure}
\begin{center}
\psfig{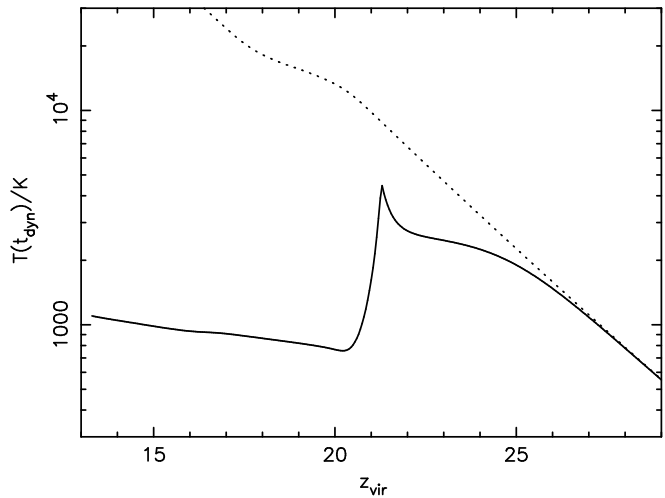}
\caption{Temperature obtained after one dynamical time versus
virialization redshift, for cosmology \LCDM.  The dotted line shows
the initial, virial temperature of the halo.} 
\label{fig:lcdm2}
\end{center}
\end{figure}

\begin{figure}
\begin{center}
\psfig{width=8.7cm,angle=270,file=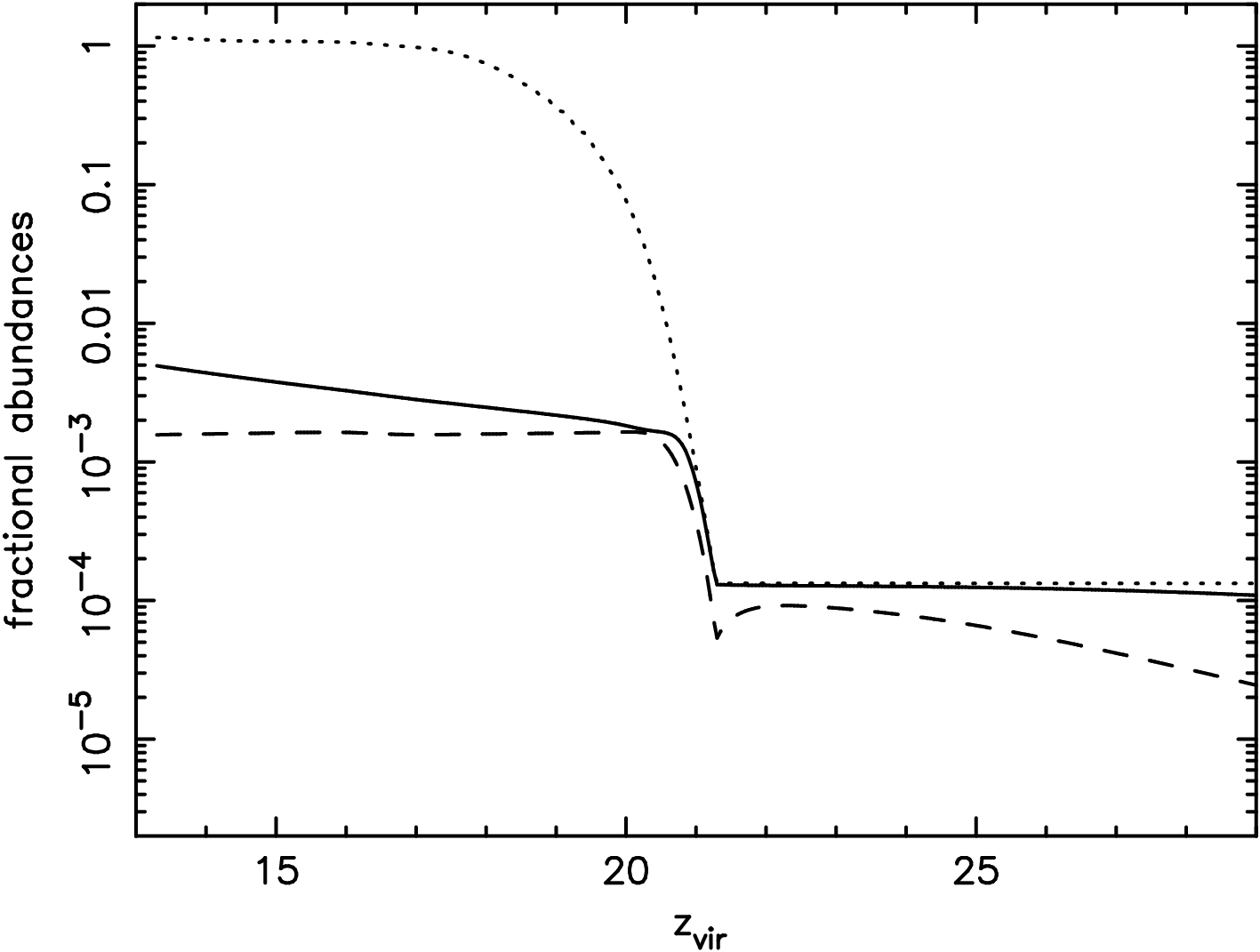}
\caption{Ionization fraction (solid line) and \htwo\ abundance (dashed
line) after one dynamical time, for cosmology \LCDM.  The dotted line
shows the initial ionization level of the halo.} 
\label{fig:lcdm3}
\end{center}
\end{figure}

This cosmology is currently the most popular CDM model.  Halos
virialize at higher redshift than in \TCDM\ but qualitatively, their
behaviour is just the same.  Figures~\ref{fig:lcdm1}--\ref{fig:lcdm3}
mimic those of Figures~\ref{fig:tcdm1}--\ref{fig:tcdm3} and show all
the same features.  In this cosmology the first halos to cool to \Tsf\
have virial temperatures of 3\,400\,K, masses of $8.9\times10^5$\Msun
and cool to 75 per cent of their virial temperature at
$z_{0.75}\approx21.8$; the second generation of higher mass halos cool
at $z\approx20.6$ and have virial temperatures and masses of 10\,600\,K
and $5.9\times10^6$\Msun.

%% file: discuss.tex
\section{Discussion}
\label{sec:discuss}

The properties of the halos corresponding to the two peaks in
Figures~\ref{fig:scdm1}, \ref{fig:tcdm1} and \ref{fig:lcdm1} are
summarized in Table~\ref{tab:halo}.

\begin{table}
\caption{Properties of the two generations of halos to form in each
of the cosmologies (corresponding to the molecular hydrogen and
electron-cooling peaks of the curves in
Figures~\ref{fig:scdm1}, \ref{fig:tcdm1} and \ref{fig:lcdm1}):
cosmological model; redshift at which the halo cools to 75 per cent of
the virial temperature; virial temperature; total mass of the halo;
baryonic mass of the halo.}
\label{tab:halo}
\begin{center}
\begin{tabular}{rcccc}
Model& $z_{0.75}$& \Tvir/K& $M_{\rm tot}/\Msun$& $M_{\rm bary}/\Msun$\\
\hline
Generation~1\\
\SCDM& 19.4& 3\,700& $1.6\times10^6$& $1.2\times10^5$\\
\TCDM& 10.8& 4\,500& $5.0\times10^6$& $9.1\times10^5$\\
\LCDM& 21.8& 3\,400& $8.9\times10^5$& $9.7\times10^4$\\
Generation~2\\
\SCDM& 18.5& 10\,800& $9.9\times10^6$& $7.6\times10^5$\\
\TCDM& 10.7& 10\,600& $2.1\times10^7$& $3.8\times10^6$\\
\LCDM& 20.6& 10\,600& $5.9\times10^6$& $6.4\times10^5$
\end{tabular}
\end{center}
\end{table}

Note that, although the redshift at which the first objects cool is
quite different in different cosmologies, their virial temperatures are
very similar, reflecting the similar chemical processes that are
occurring within them.

\subsection{The effect of cosmology}

Although it is often not stated explicitly, previous work seems to
concentrate on the \SCDM\ cosmology which is now known to be a poor
model of the Universe; the currently favoured model being \LCDM.
Coincidentally these two give similar formation redshifts for the two
generations of halos and have similar baryonic masses.  The
phenomenologically-based critical-density \TCDM\ cosmology, on the
other hand, forms its first objects at much lower redshifts, but
nevertheless still comfortably above the observed redshifts of the
most distant objects yet observed.

The baryonic masses of these first cooled halos are a factor of 10
times larger in the \TCDM\ than in the \LCDM\ cosmology---in the
former they correspond roughly to the mass of globular clusters.
Note, however, that the peaks corresponding to the first generation
halos in Figures~\ref{fig:scdm1}, \ref{fig:tcdm1} and \ref{fig:lcdm1}
are very broad so that we would expect halos with a wide range of
masses to cool almost simultaneously.

\subsection{The \htwo\ mass fraction required for collapse}

We have shown in Section~\ref{sec:scdm}, that the virial temperature
required for collapse agrees with the results from T97 and Abel \etal\
(1998).  These two papers both claim that the \htwo\ fraction required
for collapse of the first halos is $5\times10^{-4}$, a result
confirmed by Fuller \& Couchman (2000).  However, the \htwo\ fractions
shown in Figures~\ref{fig:tcdm2} and \ref{fig:lcdm2} are much lower,
about $1\times10^{-4}$.  There is no contradiction here, however.  The
Figures show the \htwo\ fraction after one dynamical time, whereas T97
give the asymptotic \htwo\ fraction at late times---for which we get a
similar value of $4.4\times10^{-4}$ for the \SCDM\ cosmology
(similarly $4.2\times10^{-4}$ for \LCDM\ and a slightly lower value of
$3.2\times10^{-4}$ for \TCDM).

\subsection{The relative roles of the two generations of halos}
\label{sec:twogen}

When talking about the first cosmological objects, we mean all those
halos that can collapse and form stars before polluted by feedback of
metals (and energy) from supernovae.  Table~\ref{tab:time} lists the
differences in collapse times for the two generations of halos in each
cosmology; these are similar to the life-time of the most massive
stars (a 15 solar mass star has a lifetime of 10\,Myr).  


The time differences in Table~\ref{tab:time} are likely to be upper
limits because the enhanced \htwo\ production will lead to more rapid
cooling at low temperatures for the second generation halos.  The
question arises as to whether, even if small objects are the first to
collapse, larger objects may collapse around them and form stars
before the first generation of supernovae explode.

The results presented in this paper are for 3\,$\sigma$ density
fluctuations.  In reality there will be fluctuations with a continuous
range of overdensities, leading to an enormous spread in collapse
times.  Note also that for the masses we are dealing with, the power
spectrum is almost flat---in this situation Press-Schecter theory
(Press \& Schechter 1974) predicts that there is as much collapsed
mass in each decade of mass as any other.

We have taken a simple model in which halos have no substructure.  In
reality, high-mass object will contain larger fluctuations on smaller
scales.  For example, the inclusion of density profiles may raise the
core density of even small halos to a point where free-fall collapse
could proceed.  In this case, we should also consider the effect of
ionizing radiation from high-mass stars; however the amount of
ionizing radiation and the degree to which this permeates the halo
will be highly-dependent upon the initial-mass-function and the
physical conditions in the star-forming region.

The halo-in-halo problem can be addressed using a statistical model of
the collapse of halos but a realistic estimate of the star-formation
time and the structure of star-forming regions requires numerical
simulations of the dynamics of the collapse as discussed in the next
section.

\begin{table}
\caption{The redshift, $z_{0.75}$ and age, $t_{0.75}$, of the universe
at the time at which the first halos in each generation have cooled to
75 per cent of the virial temperature.  The final column gives the
difference in age between the two generations.}
\label{tab:time}
\begin{center}
\begin{tabular}{rccccc}
Model& $z_{0.75}$& $t_{0.75}/$Myr& $z_{0.75}$& $t_{0.75}/$Myr& $\Delta
t$/Myr\\
\hline
& \multicolumn{2}{c}{Generation~1\hspace*{1em}}&
\multicolumn{2}{c}{Generation~2\hspace*{1em}}\\
\SCDM& 19.4& 140& 18.5& 151& 10\\
\TCDM& 10.8& 322& 10.7& 326& \ 4\\
\LCDM& 21.8& 145& 20.6& 157& 12  
\end{tabular}
\end{center}
\end{table}

\subsection{Towards more realistic models}

Our model does not attempt to follow the increasing density in objects
that do cool on less than a dynamical time.  To do so requires the use
of some form of hydrodynamics code.  Several groups have attempted
this.  Haiman, Thoul \& Loeb (1996), Omukai \& Nishi (1998) and
Nakamura \& Umemura (1999) have all simulated spherically-symmetric
collapses; Fuller \& Couchman (2000) go further and simulate a top-hat
collapse with and without substructure using a three-dimensional code.
In general, these results confirm the simple analytic predictions.

One of the most important simplifications of our model is that it
ignores substructure.  Any halo that can cool and collapse on a
dynamical time will contain smaller fluctuations of even higher
overdensity.  This will result in fragmentation of the cloud as the
collapse proceeds.  This has been investigated by Abel \etal\ (1998)
and Abel, Bryan \& Norman (2000) who have used a hierarchical grid
code to perform very high-resolution simulations of the collapse and
fragmentation of the first objects, down to scales of just 1\,pc.
They find a very filamentary structure develops with the first stars
forming from small knots at the intersection of filaments.  Only a
small fraction of the gas reaches sufficiently high densities to allow
star-formation.  If these first stars were to feed back energy into
the surrounding cloud and disrupt it, then that would suggest that the
size of the first star clusters may be much smaller than the size of
the cloud from which they formed.  However, the masses of these star
clusters in the simulations is small and may not produce many
high-mass stars.  Also, the simulations have not yet been run for long
enough to see whether the surrounding cloud will collapse before
feedback becomes effective, as we have suggested in
Section~\ref{sec:twogen}.  It should be noted that very different
results were obtained by Bromm, Coppi \& Larson (1999) using a
particle-based hydrodynamics, but with much poorer resolution.  They
found that the cloud collapsed to a rotationally-supported disk which
then broke up into very massive star clusters.  It will be a while, we
suspect, before any concensus emerges.

Fuller \& Couchman (2000) simulated a cubical region of side
25\,$h^{-1}$kpc in the SCDM cosmology with several different
realisations of the density fluctuation spectrum, using an N-body,
hydrodynamics code.  This random realisation is just what is required
to look at the relative importance of halos of different mass.  They
found that the most massive objects that collapsed within the region
did so at a wide range of redshifts, 15--30, and had a similar large
spread in mass.  Unfortunately, their simulation volume was not large
enough to sample our second-generation halos.

%% file: conc.tex
\section{Conclusion}
\label{sec:conclusions}

In this paper we have considered the cooling of gas within spherical,
virialized halos in the high-redshift Universe.  Our technique is
similar to that used by T97, but with a more up-date
cooling function and cosmological model.  In addition, we have
investigated halos with a wider range of virial temperatures, in the
range of 100\,K to 100\,000\,K.  

We have followed the abundances of Hydrogen and Helium species,
including molecular Hydrogen.  We use a more complete destruction term
for \htwo, taking into account the relative importance of the reverse
reaction H7 which other authors have ignored.  This makes a difference
of a few per cent to the final \htwo\ abundances for halos which cool
from above 9\,000\,K.

The main coolant for temperatures below 9\,000\,K is molecular
hydrogen.  Unfortunately, the cooling rate seems poorly-known and a
recent determination by Galli and Palla (1998) gives values at
temperatures below 1\,000\,K over an order of magnitude lower than
those of Lepp and Shull (1984).  Consequently the first objects to
cool in the former do so later and have much higher masses and virial
temperatures.

We follow T97 by defining clouds to have cooled if they lose 25 per
cent or more of their energy in the time that the redshift has
decreased by 25 per cent.  We obtain similar virial temperatures for
the smallest halos that can cool, but note that the \emph{first}
objects to cool in each cosmology are more massive and have higher
virial temperatures of about 4\,000\,K.  In the
\SCDM\ and \LCDM\ cosmologies the formation redshift is
$z_{0.75}\approx$20; for the \TCDM\ cosmology it is much lower,
$z_{0.75}\approx$11.

We identify a second generation of halos that cool about 10\,Myr after
the first one.  These are halos with virial temperatures in excess of
9\,000\,K for which there is a significant fraction of free electrons.
The cooling is dominated by electronic transitions at high
temperatures and is almost instantaneous, occurring on much less than
a dynamical time.  Just as significant, however, is the fact that the
residual ionization is greater than in low-mass halos and the
production of \htwo\ is much greater.  Consequently, they cool to much
lower temperatures in a dynamical time than do the first generation of
halos.

Our model suffers from three deficiencies: we consider only 3$\sigma$
density fluctuations, we ignore substructure, and we cannot follow the
collapse of halos whose cooling times are shorter than their dynamical
times.  More sophisticated studies are required to determine whether
the first or second generation halos are more important for
determining the mass of the first star-clusters, or whether they both
have a role to play.  We hope to report on these in future papers.

%% file: appendix.tex
\begin{appendix}

The appendices describes two different ways of deriving
Equation~\ref{eq:htwolong} for the rate of change of abundance of
\htwo.  We do this because nowhere in the literature does it seem to
be spelt out in detail and because our equation differs slightly from
those used previously, as described in Section~\ref{sec:h2chem}.

\section{Derivation via removal of \hminus\ and \htwoplus\ from the
reaction network}
\label{sec:hminus}

The reaction rates for destruction of \hminus\ and \htwoplus\ are much
greater than for formation or destruction of H, \hplus, \e\ and \htwo.
Hence we can regard \hminus\ and \htwoplus\ as short-lived species
that are destroyed the instant that they are produced.  This has the
advantage that we can eliminate them from the reaction network, as
described below.

Consider first \hminus.  This has two destruction channels, Reactions
H4 and H5 in Table~\ref{hmtab}, whose combined rate is
\begin{equation}
\Rhminus =\nh R_4+\nhplus R_5,
\label{eq:hminus}
\end{equation}
where $R_4$ and $R_5$ are the reaction rates as listed in the Table
(at redshifts greater than 110 photo-destruction of \hminus\ by cosmic
microwave background photons is also important, but we do not consider
such high redshifts in this paper).  Hence the fractions of \hminus\
that decay via reactions H4 and H5 are $\nh R_4/\Rhminus $ and
$\nhplus\ R_5/\Rhminus $, respectively.   Reaction H4 is always
important, whereas Reaction H5 is significant only at high ionization
levels. 

Now Reaction H3 for \hminus\ production can instead be rewritten as two
different reaction chains:
\begin{eqnarray}
{\rm H3H4} \hspace*{4.3em} 2\h  + \e &\longmapsto& \h _2 +\e
+\gamma \\
{\rm H3H5} \hspace*{2.2em} \h  + \e +\h ^+ &\longmapsto& 2{\rm
H} +\gamma
\end{eqnarray}
that occur at rates $\nh^2\nel R_3R_4/\Rhminus $ and $\nh\nel\nhplus
R_3R_5/\Rhminus $, respectively.  The first of these reaction chains
forms molecular hydrogen using electrons as catalysts whereas the second
leads to a reduction in the ionization level.

We can treat \htwoplus\ in the same manner.  Its decay channels are
Reactions~H7 and H8 which sum to a total destruction rate of
\begin{equation}
\Rhtwoplus =\nh R_7+\nel R_8.
\label{eq:htwoplus}
\end{equation}
Once again we have neglected photo-destruction which is important only
at very high redshifts.

Then Reaction 6 becomes the two reaction chains
\begin{eqnarray}
{\rm H6H7} \hspace*{4.4em} 2\h + \hplus &\longmapsto& \htwo
+ \hplus +\gamma \\
{\rm H6H8} \hspace*{2.6em} \h + \hplus + \e &\longmapsto& 2\h +\gamma
\end{eqnarray}
which are analogous to those for \hminus.  There is a second couplet
starting with \htwo:
\begin{eqnarray}
{\rm H9H7} \hspace*{4.1em} \htwo + \hplus +\h 
&\longmapsto& \htwo + \hplus +\h  \\
{\rm H9H8} \hspace*{3.7em} \htwo + \hplus + \e &\longmapsto& 3\h.
\end{eqnarray}
Of these, H9H7 is the more important at low ionization levels and will
strongly suppress destruction of \htwo\ via collisions with protons.

Using the reactions listed in Table~\ref{hmtab}, but replacing
Reactions H3 through H9 with the reaction chains derived above,
we can write down an equation for the rate of change
of \htwo\ abundance:
\begin{eqnarray}
\label{eq:htwolongagain}
\lefteqn{ \frac{d\nhtwo }{dt} = \nh ^2\left[
\nel {R_3R_4\over \Rhminus } + \nhplus {R_6R_7\over \Rhtwoplus }
\right] } \\
& & {} - \nhtwo 
\left[ \nhplus\nel{R_9 R_8\over\Rhtwoplus} + \nh R_{10} + \nel R_{11} 
\right]. \nonumber
\end{eqnarray} 

\section{Derivation using equilibrium values of \hminus\ and \htwoplus}
\label{sec:h2alt}

The second derivation uses the fact that the destruction rates for
\hminus\ and \htwoplus\ are high to derive equilibrium values for
\nhminus\ and \nhtwoplus\ which can then be eliminated from the
equations.

Consider first the equation for the rate of change of \nhminus:
\begin{equation}
{d\nhminus\over dt}=\nh \nel R_3 -
\nhminus\left[\nh R_4+\nhplus R_5\right].
\end{equation}
The destruction rates within the square brackets are very large which
means that \nhminus\ will rapidly evolve to the equilibrium
value in which creation and destruction of \nhminus\ balance and
$d\nhminus/dt\approx0$.  Then we have
\begin{equation}
\nhminus \approx { \nh \nel R_3 
\over \nh R_4+\nhplus R_5 }
\equiv { \nh \nel R_3 \over \Rhminus  }.
\end{equation}

Similarly for \nhtwoplus\ we have
\begin{eqnarray}
\nhtwoplus  &\approx& { 
\nh \nhplus R_6 + \nhtwo \nhplus R_9
\over \nh R_7+\nel R_8 } \nonumber\\
&\equiv& { \nh \nhplus R_6 + \nhtwo \nhplus R_9
\over \Rhtwoplus  }.
\end{eqnarray}

We can now write down the equation for the rate of change of \htwo,
then substitute for \nhminus\ and \nhtwoplus.
\begin{eqnarray}
\lefteqn{\frac{d\nhtwo }{dt} = 
\nh \nhminus R_4 + \nh \nhtwoplus R_7 }\\
\label{eq:htwoapp}
& & {}- \nhtwo 
\left[\nhplus R_9+\nh R_{10}+\nel R_{11}\right] \nonumber\\
&=& \nh^2\nel {R_3R_4\over \Rhminus }
   +\nh^2\nhplus {R_6R_7\over \Rhtwoplus }
   +\nh\nhtwo\nhplus {R_9R_7\over \Rhtwoplus } 
\nonumber\\
& & {}- \nhtwo 
\left[\nhplus R_9+\nh R_{10}+\nel R_{11}\right] \nonumber\\
&=& \nh^2\left[
\nel {R_3R_4\over \Rhminus } + \nhplus {R_6R_7\over \Rhtwoplus }
\right] \\
& & {} - \nhtwo 
\left[ \nhplus\nel{R_9 R_8\over\Rhtwoplus} + \nh R_{10} + \nel R_{11} 
\right]. \nonumber
\end{eqnarray}
Equation~\ref{eq:htwoapp} is identical to
Equation~\ref{eq:htwolongagain} derived above.

\end{appendix}